\documentclass[12pt,twoside]{article}   
\usepackage[super,sort,comma]{natbib}

\usepackage{fancyhdr}		

\newcommand{\captionv}[3]{\begin{center}\parbox{#1cm}{\caption[#2]{{\sf #3}}}
        \end{center}}

\usepackage[section]{placeins}   %

\usepackage{graphicx}

\makeatletter \renewcommand\@biblabel[1]{$^{#1}$} \makeatother
\newcommand{\cen}[1]{\begin{center} #1 \end{center}}

 \usepackage[mathlines,edtable]{lineno}

\usepackage{xr-hyper}
\usepackage{hyperref}
\hypersetup{ colorlinks,
    citecolor=blue,
    filecolor=blue,
    linkcolor=blue,
    urlcolor=blue
}

\usepackage{xcolor}

\definecolor{gray}{rgb}{0.6,0.6,0.6}
\definecolor{red}{rgb}{0.85,0,0}
\definecolor{green}{rgb}{0,0.85,0}
\definecolor{blue}{rgb}{0,0,0.85}
\definecolor{beige}{rgb}{0.92,0.87,0.78}

\usepackage{subfigure}
\usepackage{lscape} 
\usepackage{rotating,booktabs}
\usepackage{multirow}
\usepackage{multicol}

\newcommand{\Lb}{\mathcal{L}}

\usepackage[all]{hypcap}    

\hypersetup{
 breaklinks=true,
 pdftitle={CBCT-to-CT synthesis with a single neural network for head-and-neck, lung and breast cancer adaptive radiotherapy},    	
pdfauthor={Matteo Maspero},     			
 pdfsubject={subject},   
 pdfkeywords={CBCT},
 pdfcreator={mmaspero},   
 pdfproducer={UMC Utrecht}}

\begin{document}

\cen{\sf {\Large {\bfseries CBCT-to-CT synthesis with a single neural network for head-and-neck, lung and breast cancer adaptive radiotherapy} \\  
\vspace*{10mm}
Matteo Maspero$^{1,2}$, Antonetta C Houweling$^1$, Mark H F Savenije$^{1,2}$, Tristan C F van Heijst$^1$, Joost J C Verhoeff$^1$, Alexis N T J Kotte$^1$, Cornelis AT van den Berg$^{1,2}$} \\
$^1$Department of Radiotherapy, Division of Imaging \& Oncology, University Medical Center Utrecht, Heidelberglaan 100, 3508 GA Utrecht, The Netherlands; \\
$^2$Computational Imaging Group for MR diagnostics \& therapy, Center for Image Sciences, University Medical Center Utrecht, Heidelberglaan 100, 3508 GA Utrecht, The Netherlands
\vspace{5mm}\\
Version typeset \today\\
}

\pagenumbering{roman}
\setcounter{page}{1}
\pagestyle{plain}
Correspondence to m.maspero@umcutrecht.nl, matteo.maspero.it@gmail.com\\
Address: Department of Radiotherapy, Division of Imaging \& Oncology, University Medical Center Utrecht, Heidelberglaan 100, 3508 GA Utrecht, The Netherlands \\

\begin{abstract}
\noindent {\bf Purpose:} CBCT-based adaptive radiotherapy requires daily images for accurate dose calculations. This study investigates the feasibility of applying a single convolutional network to facilitate CBCT-to-CT synthesis for head-and-neck, lung and breast cancer patients.\\
{\bf Methods:} Ninety-nine patients diagnosed with head-and-neck, lung or breast cancer undergoing radiotherapy with CBCT-based position verification were included in this study.
CBCTs were registered to planning CTs according to clinical procedures. Three cycle-consistent generative adversarial networks (cycle-GANs) were trained in an unpaired manner on 15 patients 
per anatomical site generating synthetic-CTs (sCTs). Another network was trained with all the anatomical sites together. 
Performances of all four networks were compared and evaluated for image similarity against rescan CT (rCT). 
Clinical plans were recalculated on CT and sCT and analysed through voxel-based dose differences and $\gamma$-analysis.\\
{\bf Results:} A sCT was generated in 10 seconds. Image similarity was comparable between models trained on different anatomical sites and a single model for all sites. 
Mean dose differences $<0.5\%$ were obtained in high-dose regions. Mean gamma (2$\%$,2mm) pass-rates $>95\%$ were achieved for all sites. \\
{\bf Conclusions:}  Cycle-GAN reduced CBCT artefacts and increased HU similarity to CT, enabling sCT-based dose calculations. The speed of the network can facilitate online adaptive radiotherapy using a single network for head-and-neck, lung and breast cancer patients. \\

\end{abstract}



\tableofcontents

\newpage

\setlength{\baselineskip}{0.7cm}      

\pagenumbering{arabic}
\setcounter{page}{1}
\pagestyle{fancy}

\section{Introduction}

In modern external beam image-guided radiotherapy (IGRT), cone-beam computed tomography (CBCT) plays a crucial role in accurate patient position verification~\cite{Cho1995,Boda-Heggemann2011,Jaffray2012}.
Also, CBCT can facilitate adaptive radiotherapy (ART) by visualising daily anatomical variations~\cite{Yan1997,Wu2011}. 

CBCT image quality is inferior to that of CT in soft-tissue contrast and Hounsfield Units (HU) consistency due to the presence of artefacts~\cite{Barrett2004,Schulze2011,Remeijer2019}. 
Therefore, CBCT is not sufficient to perform accurate dose calculations\cite{Korreman2010} and patients need to be referred for a rescan CT (rCT) whenever anatomical differences are noted between daily images and planning CT~\cite{Ramella2017}.
However, scheduling and acquiring a rCT adds logistic complexity and patient burden to the treatment.
On the contrary, with ART these issues can be addressed by
exploiting the daily CBCT images to reduce set-up errors and eliminate the need for an rCT~\cite{Wu2011}. 
A prerequisite for online ART is that the CBCT quality and HU accuracy is sufficient to enable dose calculation.

Considerable literature has recently emerged proposing to correct CBCT imaging artefacts and increase image intensity consistency using:
look-up table-based approaches~\cite{Dunlop2015,Kurz2015}, deformable imaging registration (DIR) of the planning CT to the daily anatomy on CBCT~\cite{Zhen2012,Veiga2014,Veiga2016} and model- or Monte Carlo-based methods for scatter estimation and correction~\cite{Jarry2006,Bootsma2014,Zhao2016}. 
Specifically, DIR enabled accurate dose calculations for head-and-neck (HN)~\cite{Peroni2012} but obtained lower dose accuracy in more complex anatomical changes such as lung~\cite{Veiga2016} and pelvis~\cite{Kurz2016,Giacometti2019}.
Also, Monte Carlo-based methods were suitable for ART~\cite{Niu2010,Park2015,Kurz2016}. 
These techniques can be deployed on a time scale of minutes, which is not acceptable when aiming to use CBCT images for daily online dose evaluation or online pre-treatment adaptation.

Recently, deep learning has been proposed for fast CBCT artefact correction~\cite{Kida2018,Xie2018,Maier2019,Hansen2018,Liang2019,Kurz2019}.
Deep learning is a branch of artificial intelligence and machine learning that involves the use of neural networks to generate a hierarchical representation of the input data
to achieve a specific task without the need of hand-engineered features~\cite{Meyer2018,Sahiner2019}.
Deep learning has shown promising results solving image-to-image translation problems within seconds~\cite{Han2017,Maspero2018}. 
In this sense, previous work demonstrated the use of a two-dimensional (2D) U-net to improve CBCT image quality~\cite{Kida2018,Xie2018,Maier2019}.
Moreover, it has been shown that converting CBCT with deep learning resulted in accurate dose calculation for prostate cancer patients~\cite{Hansen2018,Kurz2019} and HN cancer patients~\cite{Liang2019,Harms2019}.

In this study, we investigate whether CBCTs converted with convolutional networks may be used as a surrogate of the daily anatomy for dose calculations.
We employ a network trained in an unpaired manner to convert CBCT-to-CT of HN, lung, or breast cancer patients investigating whether a single network can generalise for the three anatomical sites. A single network trained for all the anatomical sites was compared to three networks trained per anatomical site.
Performances of all four networks were compared and evaluated for image similarity and dose calculation accuracy between CT and rCT.

\section{Material and methods}
\subsection{Imaging protocols}
\label{sec:data} 
Ninety-nine patients diagnosed with HN (33), lung (33) or breast (33) cancer undergoing radiotherapy were retrospectively included in this study. 
Irradiations were performed between May 2016 and February 2019 on Agility linacs (Elekta AB, Sweden) with CBCT-based pre-treatment position verification.

An rCT was acquired in case anatomical variations were noted on the CBCT. We included at least fourteen patients with rCT per site. 

The (r)CTs were acquired on a Brilliance Big Bore (Philips Healthcare, Ohio, USA); CBCTs were acquired using X-ray volumetric imaging (XVI, v5.0.2b72 Elekta AB, Sweden) system.
Table~\ref{tab:imagparam} reports the imaging protocols for CT, rescan CT (rCT) and CBCT for all the patients included in the study.
CBCTs were acquired with 0.25~rotation/s gantry speed and 5.5 frames/s.
All the CBCTs were acquired with a 200$^{\circ}$-arc utilising an empty filter cassette (F0) in combination with a centred detector panel (S position, maximum FOV=27x27~cm$^2$). The field-of-view (FOV) was in four cases (elective lymph-nodes irradiations or double-sided irradiation for breast and HN patients) enlarged to a maximum of 41x41 cm$^2$ using a shifted detector panel (M position) to accommodate the CTV in the CBCT FOV.

\begin{table}[!ht]
\renewcommand{\tabcolsep}{0.1cm}
\vspace{-15pt}
    \centering
\captionv{16}{Overview of the imaging protocols}{Overview of CT (including also rescan (r)CT) and CBCT imaging protocols in terms of field-of-view (FOV), acquisition matrix (Acq matrix), resolution (Res), tube voltage (kVp), exposure (ms) and current (mA). For exposure and current, the mean value ($\pm\sigma$) was reported along with the range.
 } \label{tab:imagparam}
 \small{
\begin{tabular}{|c|c|c|c|c|c|c|c|}
 \hline
\multirow{ 2}{*}{\textbf{Modality}} & \multirow{ 2}{*}{\textbf{Site}} & \textbf{FOV}$^{a}$ & \multirow{ 2}{*}{\textbf{Acq matrix}$^{a}$} & \textbf{Res}$^{a}$ & \textbf{Voltage}$^{b}$ & \textbf{Exposure}$^{b}$ & \textbf{Current}$^{b}$  \\ 
			 & 			   & [cm$^3$] & & [mm$^3$] &  [kVp] & [ms] &  [mA]   \\
\hline   
			 & \multirow{ 3}{*}{Head-and-neck}  & 43-70 & 512 & 0.83-1.37 & \multirow{ 3}{*}{120} & \multirow{1.5}{*}{983$\pm$65} & \multirow{1.5}{*}{159$\pm$50} \\
			 &								& 43-70 & 512 & 0.83-1.37	&			 & \multirow{1.5}{*}{923-1090} & \multirow{1.5}{*}{47-271} \\
			 &								& 30-111 & 101-535 & 2-3	&			 & & \\        \cline{2-8} 
\multirow{ 3}{*}{\textbf{(r)CT}} & \multirow{ 3}{*}{Breast}   & 47-70 & 512 & 0.92-1.37 & \multirow{ 3}{*}{120} &  \multirow{1.5}{*}{1050$\pm$109} &  \multirow{1.5}{*}{63$\pm$37} \\
			 &							  & 47-70 & 512 & 0.92-1.37	&			 & \multirow{1.5}{*}{923-1332} & \multirow{1.5}{*}{31-271} \\
			 &							  & 31-120 & 103-400 & 2-3	&			 & & \\ \cline{2-8}			 
			 & \multirow{ 3}{*}{Lung}  	  & 29-70 & 512 & 0.57-1.37 & \multirow{ 3}{*}{120} &  \multirow{1.5}{*}{3886$\pm$3095} &  \multirow{1.5}{*}{98$\pm$66} \\
			 &							  & 29-70 & 512 & 0.57-1.37	&			 & \multirow{1.5}{*}{500-10091} & \multirow{1.5}{*}{30-271} \\
			 &							  & 23-220 & 76-660 & 1-3	&			 & & \\ 			 	
\hline
			 & \multirow{ 3}{*}{Head-and-neck}  & 27 & 135-270 & 1-2 & \multirow{ 3}{*}{100$^c$} & \multirow{1.5}{*}{11$\pm$5} & \multirow{1.5}{*}{14$\pm$3} \\
			 &								& 27 & 135-270 & 1-2	&			 & \multirow{1.5}{*}{10-40} & \multirow{1.5}{*}{10-20} \\
			 &								& 13-53 & 126-526 & 1-2	&			 & & \\ \cline{2-8}
 \multirow{ 3}{*}{\textbf{CBCT}}& \multirow{ 3}{*}{Breast}  & 27-41 & 270-540 & 0.5-1 & \multirow{ 3}{*}{120} &  \multirow{1.5}{*}{33$\pm$2} &  \multirow{1.5}{*}{17$\pm$2} \\
			 &											  & 27-41 & 270-540 & 0.5-1	&			 & \multirow{1.5}{*}{32-40} & \multirow{1.5}{*}{16-20} \\
			 &											  & 26-53 & 262-526 & 0.5-1	&			 & & \\ \cline{2-8}			 
			 & \multirow{ 3}{*}{Lung}  & 27 & 270 & 1-2 & \multirow{ 3}{*}{120$^{d}$} &  \multirow{1.5}{*}{31$\pm6$} &  \multirow{1.5}{*}{20$\pm$1} \\
			 &						   & 27 & 270 & 1-2	&			 & \multirow{1.5}{*}{10-40} & \multirow{1.5}{*}{16-25} \\
			 &						   & 26-53 & 128-528 & 1-2	&			 & & \\ 	
\hline
 \end{tabular}
\footnotesize{
 $^{a}$Expressed in RL, AP, FH directions; the range is reported in terms of min-max.\\
 $^{b}$ Reported in terms of mean value and range=min-max.\\
 $^{c}$Except for H18 and H20 where kVP was 120.\\
 $^{d}$Except for L11, L22 and L24 where kVP was 100.\\
} 
  }
\end{table}

Imaging frequency of CBCT followed the extended non-action limit protocol \cite{deBoer2001}: online corrections (action level 0~mm) were applied in the case of partial or ablative breast irradiation, and offline long (N=3, P=5) and short (N=2, P=3) scheme were applied for irradiations having $>20$ and $<20$ fractions, respectively. Imaging frequency may have been increased after consultation between a medical physicist and a radiotherapist on a single patient-basis in case large inter-fraction motions were observed in the initial fractions or whenever RT technicians reported difficulties in reproducing the planning position. 

CBCTs were translated to apply clinical set-up corrections and resampled to the planning CT within the X-ray volumetric imaging (XVI, v5.0.2b72 Elekta AB, Sweden) system.
Registrations were estimated within a clip-box including the CTV based on bone rigid (translation and rotation) matching \cite{Borgefors1988}. 
For the breast patients treated with local RT followed by a sequential boost, a dual rigid registration was performed based first on bone matching followed by grey level (soft-tissue) matching \cite{Hristov1996,Roche1998}. The centre of rotation was assigned as the centre of the PTV.

\subsection{CBCT-to-CT synthesis}

\subsubsection{Image pre-processing} 
\label{sec:pre-proc}
Before supplying images to the network, CT and CBCT were cropped to the size of the CBCT FOV after identifying the so-called ``Mask$_{\textrm{\tiny CBCT}}$'' according to the following steps.
CBCTs were thresholded at -999.9, obtaining a binary mask. In each transverse slice containing the binary mask, morphological closure was performed, and
the smallest bounding box containing the mask was found. The biggest circle contained in the bounding box was searched starting from a radius of 26.9~cm and iteratively increasing its size. The circle was propagated for all the slices obtaining Mask$_{\textrm{\tiny CBCT}}$. CT and CBCT were cropped in the bounding box containing Mask$_{\textrm{\tiny CBCT}}$.

In addition to cropping, voxel intensity of CT and CBCT were clipped within the interval [$-$1000;3071]~HU and image intensity was linearly rescaled to 16-bit.

\begin{figure}[!hbt]
   \begin{center}
\includegraphics[width=0.95\linewidth]{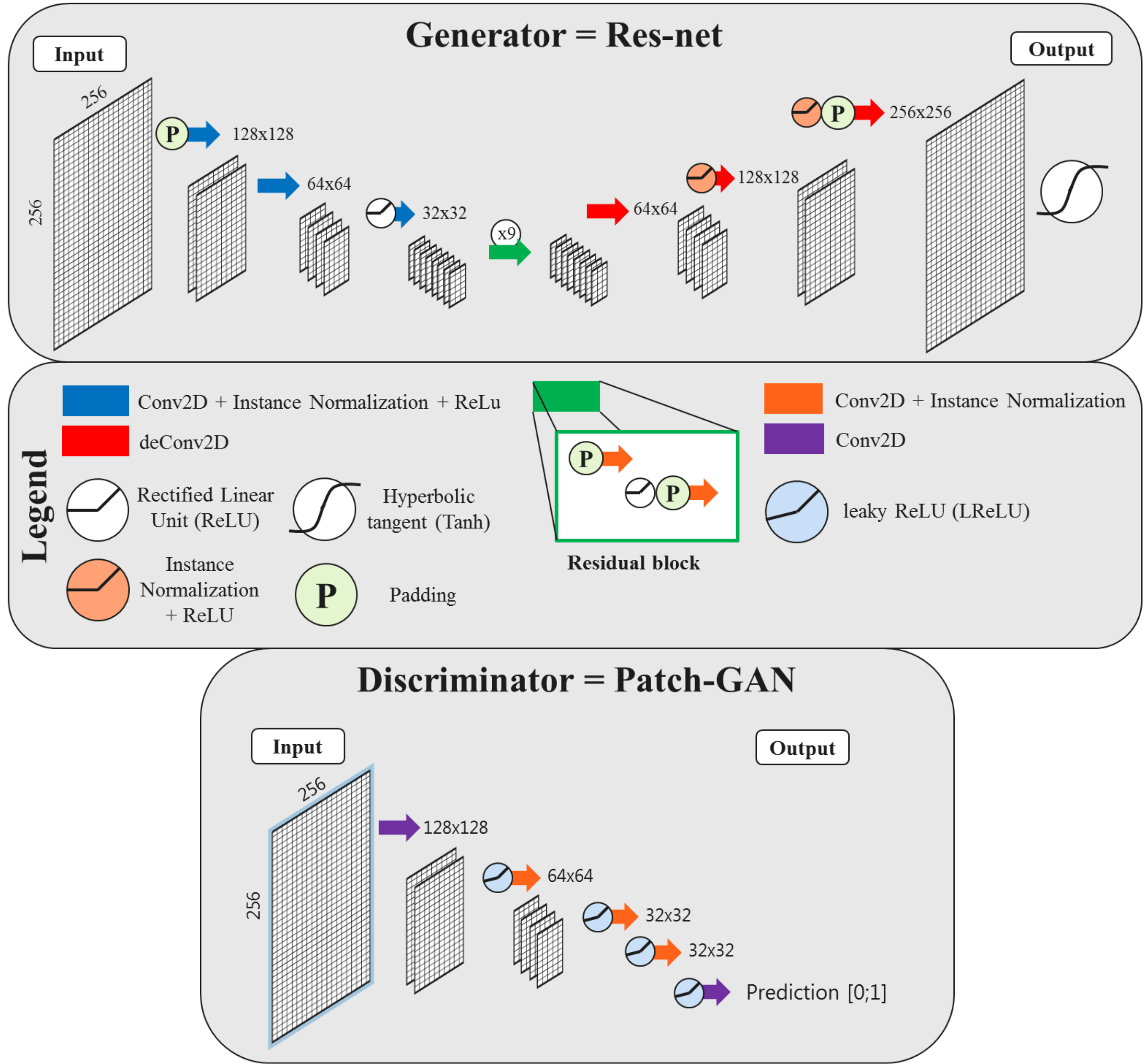}
\captionv{16}{}{Architecture of the nine-block residual network used as a generator (top) and of the convolutional network called Patch-GAN used as a discriminator. The size of the images is numerically reported, except for the residual block, where it remains stable. Note that the nine-blocks are omitted in the schematic. Each of the filters had stride two, kernel size four; leaky rectilinear rectifier unit had a scalar multiplier of 0.2, padding was applied in reflect mode.
   \label{fig:NetArch}
   }
   \end{center}
\end{figure}

\subsubsection{Network architecture and training}
\label{sec:train}

To generate CT from CBCT, a 2D cycle-generative adversarial network (cycle-GAN) was adopted~\cite{Zhu2017unp}. 
Cycle-GANs enable unpaired training, which, compared to paired training, makes the network less sensitive to residual mismatch of CT and CBCT~\cite{Wolterink2017}.

The network consisted of two cycles called ``forward'' and ``backwards'' during which GANs generated CT from CBCT and vice-versa.
Moreover, so-called ``cycle-consistency'' was enforced with an $\Lb_1$-norm such that after converting from CBCT to CT and vice-versa, the original image should be obtained.
The architecture, based on the cycle-GAN provided by Zhou et al.~\cite{Zhu2017unp}, was implemented in Tensorflow (v1.3.0). 
Nine-blocks residual networks~\cite{Johson2016} were employed as generators and Patch-GANs~\cite{Isola2017} as discriminators (Figure~\ref{fig:NetArch}). Stochastic gradient descendent was used applying an Adam solver \cite{Kingma2014} with learning rate~=~0.0002, momentum parameters $\beta_1$~=~0.5 and $\beta_2$~=~0.999.
Instance normalisation \cite{Ulyanov2016} was employed with a batch size of 1. The weights of the network were randomly initialised from $\mathcal{N}(0,\,0.02)$.
Weight optimisation was performed as in Goodfellow et al.~\cite{Goodfellow2014} alternating between one gradient descendent step on the discriminator network
and one step on the generator network after having performed a forward and backward cycle. A structured loss function GAN$+\lambda \cdot\Lb_1$+cycle-consistency with $\lambda$~=~25 was adopted.
The original implementation of the network by Zhou et al~\cite{Zhu2017unp}\footnote{\url{https://github.com/xhujoy/CycleGAN-tensorflow}} was modified to accommodate 16-bit grey-scale images with a size of 256x256.

To train, validate and test the network, patients were split into three datasets: 15 patient per site for training, 8 for validation and the remaining 10 for test.
The validation set was used to aid hyperparameter optimisation and to determine at which iteration the training could be stopped to avoid over-fitting (early-stopping), while the test set was used to evaluate the performance of the network.
To investigate the impact on dose calculation of a different CT, the patients included in the test set were selected among the patients with an rCT and with CBCT and rCT acquired with minimal time differences.
Patients' demographics were controlled to ensure data balancing in terms of the number of patients in the three sets. Also, we inspected the ratio of male/female, distribution age, tumour staging distribution and linac on which CBCT were acquired (Supplementary Material).

Training of the cycle-GAN was performed in the transverse plane, and for each iteration, random CT, and CBCT slices of different patients were supplied.
Three networks were trained separately on each anatomical site.  
Another network was trained on all anatomical sites to investigate whether a single network may generalise for all anatomical sites.
The networks were trained for 200 epochs on a Tesla P100 (16 Gb, NVIDIA, California, USA) graphical processing unit (GPU) with batch size one and image pool of 1000 images.
Data augmentation was applied during training by flipping the images left and right and randomly cropping of 30x30 voxels after having bi-linearly resampled the images to 286x286 voxels in Mask$_{\textrm{\tiny CBCT}}$. 
Early stopping was applied selecting the earliest epoch for which average $\Lb_1$ in the body contour over the pts of the validation test was the lowest and within one $\sigma$ from $\Lb_1$ calculated at every 10 epoch (a total of 20 models were stored, one each 10 epochs). 

The total amount of slices utilised during training was reported to verify data balancing among sites: 3668 transverse slices acquired in a minimum of four different linacs were used for training, composed by 1606, 1046 and 1016 slices from HN, lung and breast cancer patients, respectively.

\subsubsection{Image post-processing}
\label{sec:post-proc}

First, the trained model was applied within Mask$_{\textrm{\tiny CBCT}}$ to the pre-processed CBCTs (as described in \ref{sec:pre-proc}) obtaining 16-bit images. Then, the HU intensity range of [-1000;3071] was restored with a linear rescaling obtaining the so-called CBCT$_{\textrm{\small conv}}$.
CBCT$_{\textrm{\small conv}}$ were bi-linearly resampled from a matrix size of 256x256 to the original CBCT resolution.

\begin{figure}[!ht]
   \begin{center}
   \vspace{10pt}
\includegraphics[trim=1 5 10 0,clip,width=16.2cm]{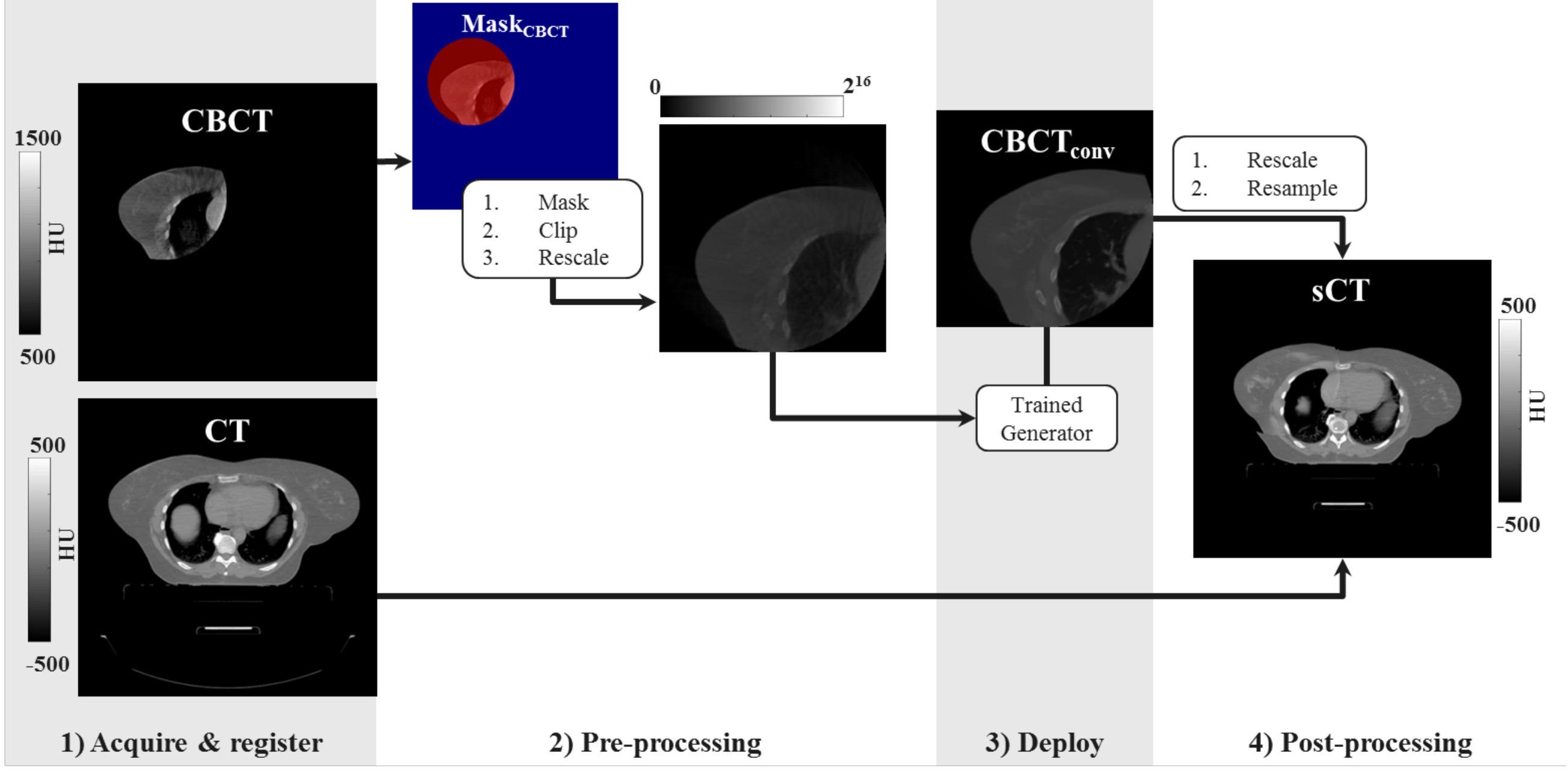}
\vspace{-15pt}
\captionv{16}{}{Schematic of the image workflow for a 2D transverse slice of a breast cancer patient. After image acquisition, registration (1) and pre-processing (2) the trained network is deployed producing converted CBCT (CBCT$_{\textrm{\small conv}}$, 3) which substituted the original CT within Mask$_{\textrm{\tiny CBCT}}$ obtaining the so-called synthetic CT (sCT).
   \label{fig:DataPip}
 }
    \end{center}
\end{figure}

To generate images for the full CT FOV, the CBCT$_{\textrm{\small conv}}$ was substituted in the original CT within the Mask$_{\textrm{\tiny CBCT}}$.
The image obtained combining CT and CBCT$_{\textrm{\small conv}}$ will be referred to as synthetic-CT (sCT) (Figure~\ref{fig:DataPip}).

\subsection{Evaluation}
\label{subsec:eval}
Image evaluation in terms of similarity between sCT and rCT was performed to assess whether the single network trained with all the anatomical sites was comparable to the three networks trained per anatomical site.
If performances were comparable, the single network was considered to assess the appropriateness of the CBCT conversion with the trained cycle-GAN on the test set with an image and a dose comparison.

\subsubsection{Image comparison}
Similarities between the image intensity of sCT, CBCT, CT and rCT were calculated within Mask$_{\textrm{\tiny CBCT}}$ in terms of mean absolute error (MAE) and mean error (ME) as proposed by Liang et al.~\cite{Liang2019}. Rescan CT was considered as ground truth, and the metrics were calculated in terms of mean $\pm1\sigma$ and range. Wilcoxon signed-rank tests were conducted between sCT/rCT and CT/rCT for MAE.
Additional metrics are reported in Supplementary Material. 

\subsubsection{Dose comparison}
For the patients in the test sets, clinical plans were recalculated on CT, sCT and rCT images in Monaco (v 5.11.02, Elekta AB, Sweden) using a Monte Carlo algorithm on a grid of 3~mm$^3$
with 5\% and 3\% statistical uncertainty for volumetric modulated arc therapy (VMAT) and intensity-modulated radiotherapy (IMRT) plans, respectively.
Clinical contours, delineated by a radiation oncologist on the planning CT, were rigidly transferred to the sCT and rCT except for the body contour, which was automatically re-delineated. These contours were considered as volumes of interest (VOIs).

Dose distributions were analysed through relative dose differences ($\textrm{DD}_{\textrm{sCT}}=\frac{\textrm{sCT}-\textrm{rCT}}{\textrm{rCT}}$ and 
$\textrm{DD}_{\textrm{CT}}=\frac{\textrm{CT}-\textrm{rCT}}{\textrm{rCT}}$) in the high dose region (dose $>90\%$ of the prescribed dose).
Also, 3D $\gamma$-analysis~\cite{Low2010} with 3\%,3mm and 2\%,2mm criteria relative to dose on rCT within regions of 50\% prescription dose were performed.
For all dose comparisons, a 15~mm cropping in the proximity of body contour was performed to take account of dose build-up in the proximity of the skin~\cite{Maspero2018REct}. 

To investigate the impact of dose difference within VOIs, analysis of dose-volume histogram (DVH) points was performed
on sCT and rCT. The DVH points analysed were the maximum dose and mean dose. OARs were considered for such analysis when they were present in at least four of the patients for each anatomical site: submandibular and parotid glands, spinal cord, larynx and brain stem for head-and-neck patients; lungs, heart, oesophagus, humerus and spinal cord for breast patients; lungs, heart, oesophagus, spinal cord and trachea for lung patients. 

\section{Results}

\subsection{Network}
Cycle-GANs required about eight days and five hours training on a GPU Tesla P100 (NVIDIA Corporation) employing 200 epochs with 3668
slices. As a result of the early stopping investigation, after inspecting the $\Lb_1$ loss function on the training and validation set, we opted for utilising 160000 ($\sim$100 epochs), 180000 ($\sim$160 epochs), 180000 ($\sim$170 epochs) iterations for the network trained on HN, breast and lung dataset, respectively, and 360000 iterations ($\sim$100 epochs) for the network trained on all the three sites combined. 
Generating sCT required $<$10~s for an entire CBCT volume ($\sim$70 slices) on GPU and about 40~s on CPU.

\subsection{Image comparison}
The time between rCT and CBCT in the test set was on average ($\pm\sigma$ [min; max]) $1\pm3$ [0;8] days and $29\pm11$ [8;67] days between CT and CBCT.
The increased time between CT and CBCT may result in larger differences when comparing rCT vs CT compared to sCT vs rCT.
Similarity metrics over the test patients are reported in Table~\ref{tab:res_sim}.

\begin{table}[htbp]
\begin{center}
\captionv{16}{}{Overview of the image comparison. Image comparison calculated as mean ($\pm1\sigma$) and range ([min;max]) of the test dataset (30 patients) compared to the reference dataset in terms of mean absolute error (MAE) and mean error (ME) between the Test image minus the Ref image.
 } \label{tab:res_sim}
\vspace*{2ex}
\small{\begin{tabular} {|c|c| cc| cc| cc| }
\hline
\multicolumn{2}{|c|}{\textbf{Site}} & \multicolumn{2}{c|}{\textbf{Head-and-Neck}} & \multicolumn{2}{c|}{\textbf{Breast}} & \multicolumn{2}{c|}{\textbf{Lung}}  \\ 
\hline
 \multirow{ 2}{*}{\textbf{Test}} & \multirow{ 2}{*}{\textbf{Ref}}  & \textbf{MAE} &  \textbf{ME} & \textbf{MAE} &  \textbf{ME}  & \textbf{MAE} &  \textbf{ME}  \\
			 & &  [HU] & [HU]&  [HU] & [HU]&  [HU] & [HU]\\ 
\hline
 \multirow{ 2}{*}{CBCT}   	& \multirow{ 2}{*}{rCT} &   195$\pm$20 & -122$\pm$33  &   152$\pm$40 & 71$\pm$37  &  219$\pm$44 & 153$\pm$48  \\ 
	 & 		      			 		    		    &[160;230]   & [-183;-71]  &[98;213]   & [7;115] &[133;280]   & [94;230]\\  
\hline
 \multirow{ 1}{*}{sCT single} & \multirow{ 2}{*}{rCT}  & 53$\pm$12    & -3$\pm$7  & 66$\pm$18   & -6$\pm$13 &83$\pm$10   & -2$\pm$11 \\
\multirow{ 1}{*}{network$^{a}$}   						& 			 	        & [37;77]    & [-15;10]  & [41;95]     & [-24;13]  & [72;104]     & [-25;10]  \\  
\hline
 \multirow{ 1}{*}{sCT separate} & \multirow{ 2}{*}{rCT}  & 51$\pm$12   & -6$\pm$6  & 67$\pm$18   & -5$\pm$11 &86$\pm$9   & -5$\pm$14 \\
\multirow{ 1}{*}{networks$^{b}$}   						& 			 	        & [35;74]     & [-16;4]  & [41;98]     & [-18;14]  & [73;105]     & [-28;10]  \\  
\hline
 \multirow{ 2}{*}{CT}	& \multirow{ 2}{*}{rCT} &   63$\pm$17  & -18$\pm$15   &  63$\pm$24  & 8$\pm$20 &   94$\pm$23  & 9$\pm$22  \\ 
		    &                                   &[-40;90]     & [-46;3]  & [40;115]   & [-14;54]   &[68;146]     & [-33;36]  \\ 
\hline
\end{tabular} }\\
\footnotesize{
$^{a}$ sCT obtained from a single network trained on all the anatomical sites.\\
 $^{b}$ sCT obtained from three different networks trained on each anatomical site.}
\end{center}
\end{table}

\textit{Generic network vs site-specific networks}\\
No statistically significant differences (p$>$0.35) were found between networks trained per separate anatomical site and the single network trained with all three anatomical sites.
This justifies the use of the single model trained for all the anatomical sites for assessing the accuracy of HU and for the dose comparison.

\textit{Accuracy of HU}\\
One can notice that similarity increased between sCT and rCT compared to CBCT and rCT; e.g. MAE decreased
from 195$\pm$20 (CBCT/rCT) to 53$\pm$12~HU (sCT/rCT) for HN.
All the similarity metrics calculated between sCT/rCT and CT/rCT can be considered equivalent to the metrics calculated between CT and rCT, with no
significant differences (p$>$0.14) for all the three anatomical sites.
The mean MAE and range for sCT/rCT were smaller than for CT/rCT due to the reduced time between sCT/rCT, which resulted in less anatomical differences.

\begin{figure}[ht]
\begin{center}
   \vspace{-5pt}
    \begin{subfigure}{} 
  \includegraphics[trim=115 120 20 110,clip,width=\linewidth]{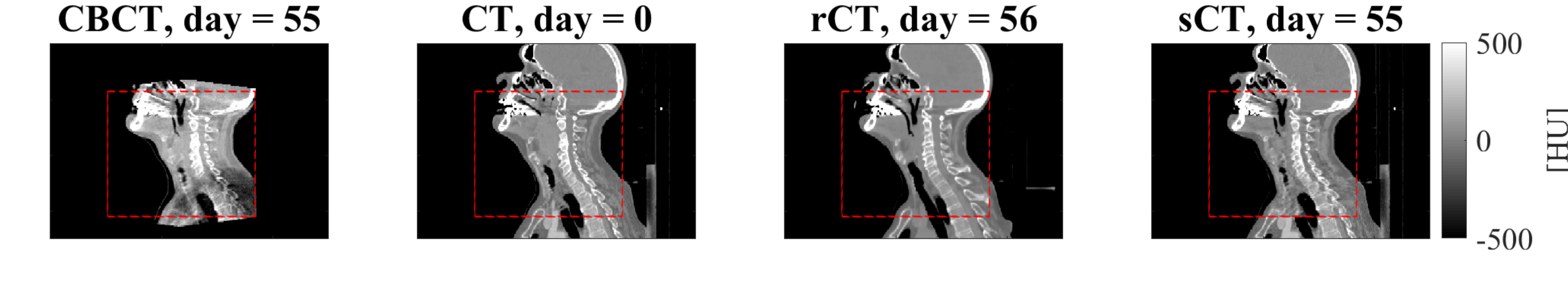} 
   \end{subfigure}   
      \vspace{-25pt}
     \begin{subfigure}{}
    \includegraphics[trim=115 140 20 110,clip,width=\linewidth]{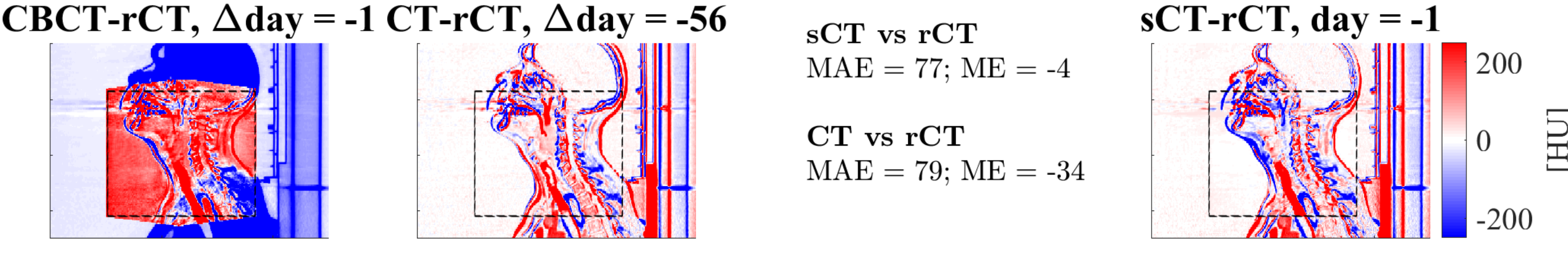}
    \end{subfigure} 
       \vspace{-25pt}
    \begin{subfigure}{} 
      \includegraphics[trim=115 140 20 100,clip,width=\linewidth]{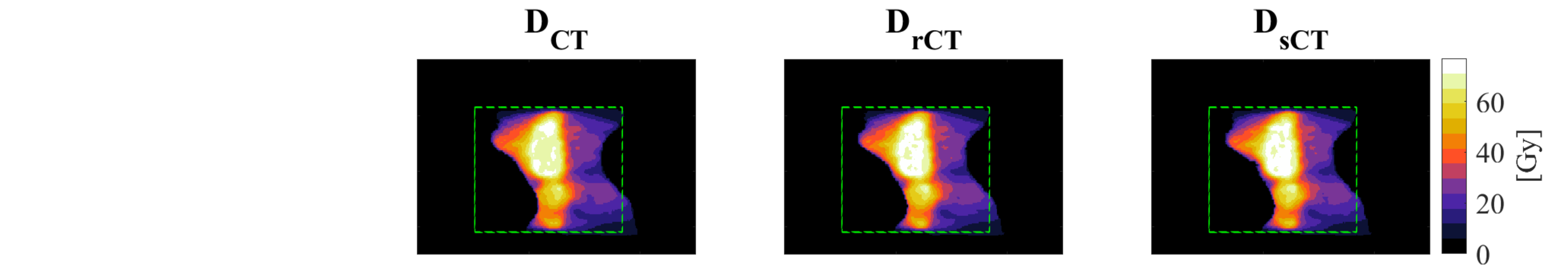}
    \end{subfigure} 
       \vspace{-25pt}
    \begin{subfigure}{}
     \includegraphics[trim=5 0 10 0,clip,width=0.9\linewidth]{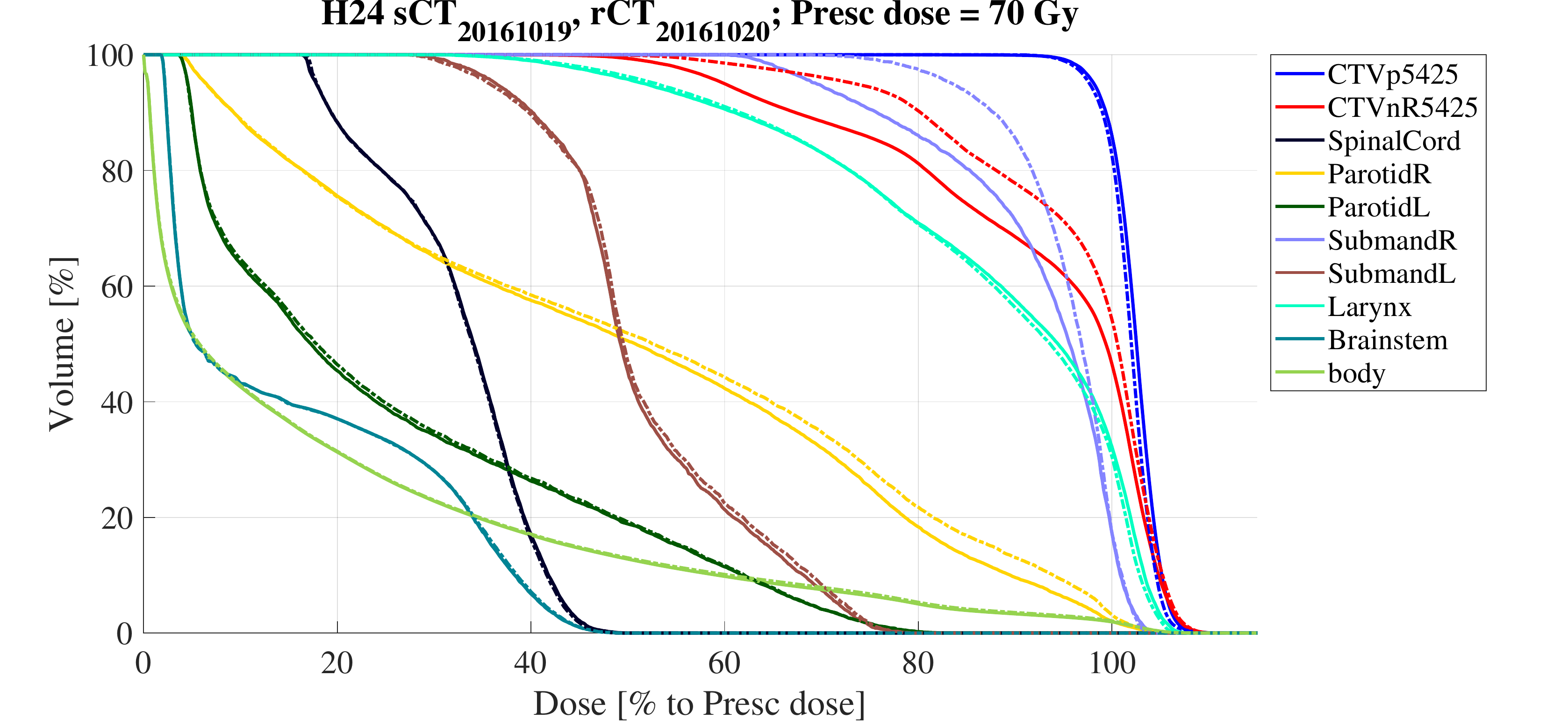} 
    \end{subfigure} 
\end{center}
\captionv{16}{}{Sagittal views for the head-and-neck cancer patient H\thect\ of: (1$^{\textrm{st}}$ row) CBCT (1$^{\textrm{st}}$ column), CT (2$^{\textrm{nd}}$ column), rescan CT (rCT, 3$^{\textrm{rd}}$ column) and synthetic CT (sCT, 4$^{\textrm{th}}$ column), along with (2$^{\textrm{nd}}$ row) the respective difference to rCT, the doses (3$^{\textrm{rd}}$ row). 
 The red, black, or green dotted rectangles indicate the position of Mask$_{\textrm{\tiny CBCT}}$. The days refer to the acquisition date of the rCT. In the 4$^{\textrm{th}}$ row, the DVH is shown for target and OARs of sCT (continuous lines) and rCT (dashed lines).
   \label{fig:Ptexample_Hn} 
 }
\end{figure}

\begin{figure}[!htbp]
\begin{center}
   \vspace{-5pt}
    \begin{subfigure}{} 
  \includegraphics[trim=115 145 20 115,clip,width=\linewidth]{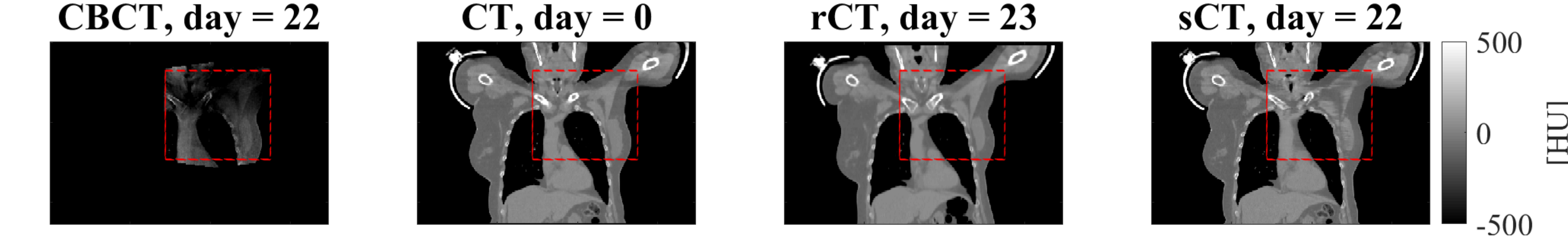} 
   \end{subfigure}   
      \vspace{-15pt}
     \begin{subfigure}{}
    \includegraphics[trim=115 145 20 115,clip,width=\linewidth]{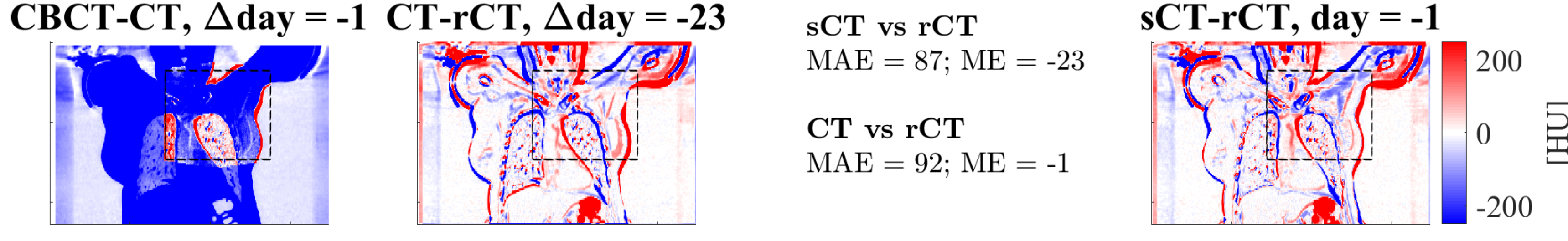}
    \end{subfigure} 
          \vspace{-15pt}
    \begin{subfigure}{} 
      \includegraphics[trim=115 150 20 110,clip,width=\linewidth]{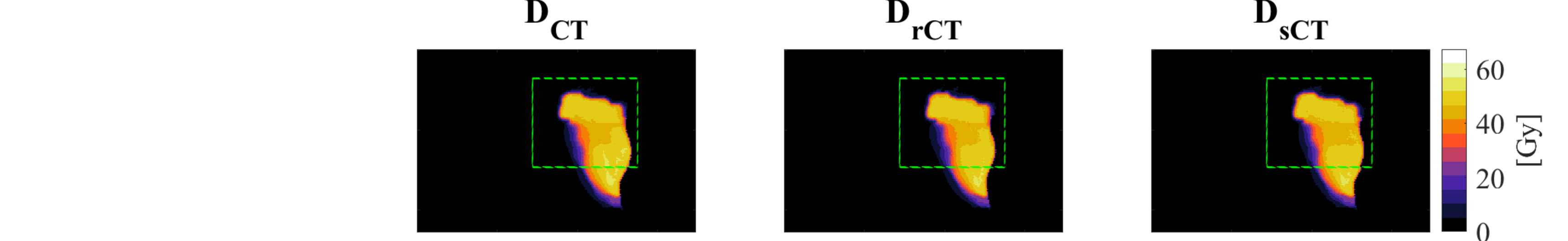}
    \end{subfigure} 
          \vspace{-15pt}
    \begin{subfigure}{}
     \includegraphics[trim=5 0 10 0,clip,width=0.9\linewidth]{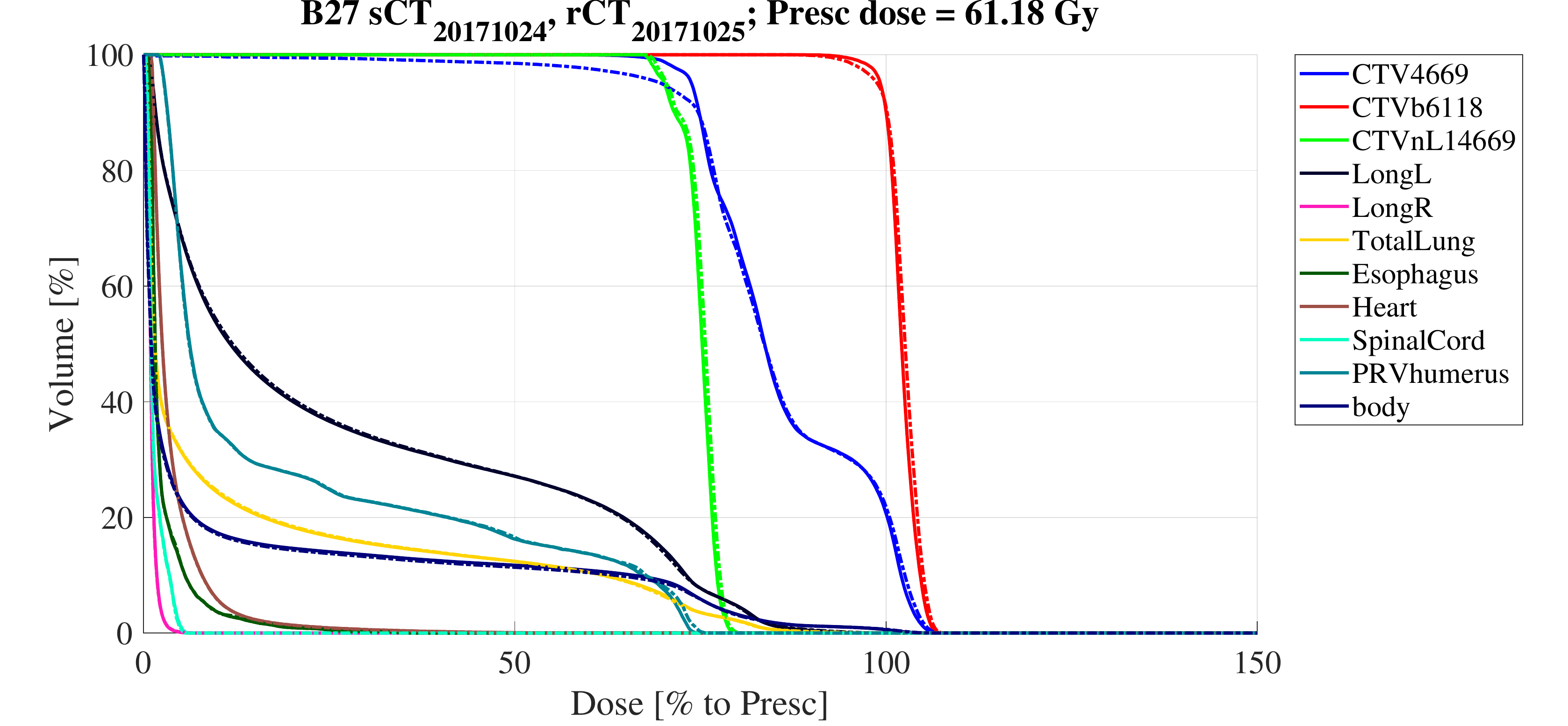} 
    \end{subfigure}     
\end{center}
    \vspace{-20pt}
\captionv{16}{}{Coronal views for the breast cancer patient B\thect\ of: (1$^{\textrm{st}}$ row) CBCT (1$^{\textrm{st}}$ column), CT (2$^{\textrm{nd}}$ column), rescan CT (rCT, 3$^{\textrm{rd}}$ column) and synthetic CT (sCT, 4$^{\textrm{th}}$ column), along with (2$^{\textrm{nd}}$ row) the respective difference to rCT, the doses (3$^{\textrm{rd}}$ row). 
 The red, black, or green dotted rectangles indicate the position of Mask$_{\textrm{\tiny CBCT}}$. The days refer to the acquisition date of the rCT. In the 4$^{\textrm{th}}$ rows, the DVH is shown for target and OARs of sCT (continuous lines) and rCT (dashed lines).
   \label{fig:Ptexample_Br} 
 }
\end{figure}

\begin{figure}[ht]
\begin{center}
   \vspace{-25pt}
 \begin{subfigure}{} 
  \includegraphics[trim=115 105 20 85,clip,width=\linewidth]{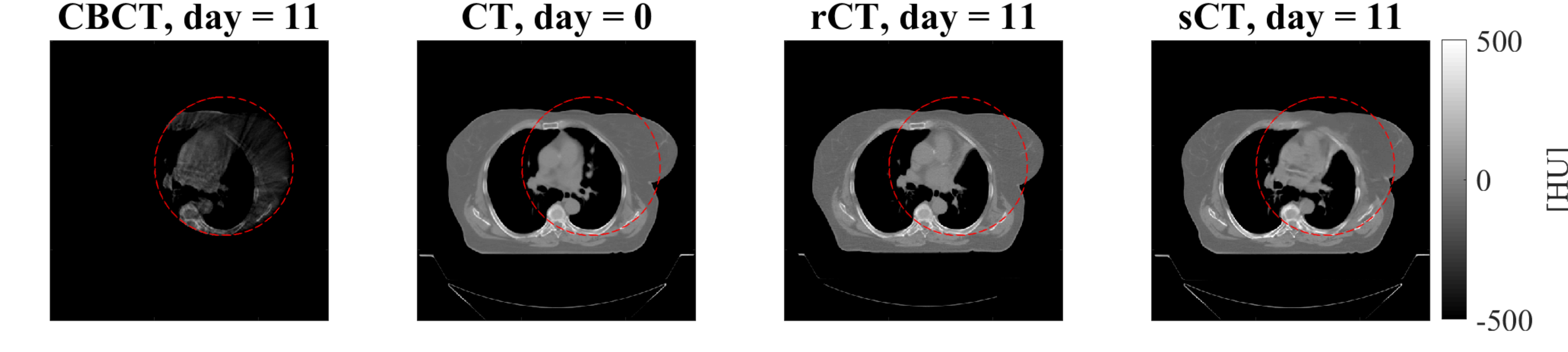} 
   \end{subfigure}         \vspace{-25pt}
     \begin{subfigure}{}
    \includegraphics[trim=115 105 20 85,clip,width=\linewidth]{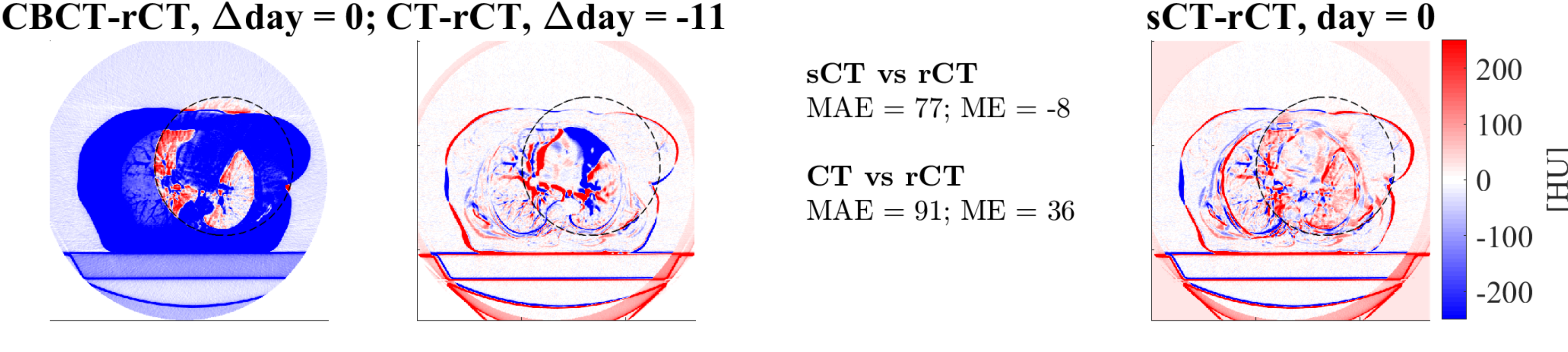}
    \end{subfigure}       \vspace{-25pt}
    \begin{subfigure}{} 
      \includegraphics[trim=115 105 20 75,clip,width=\linewidth]{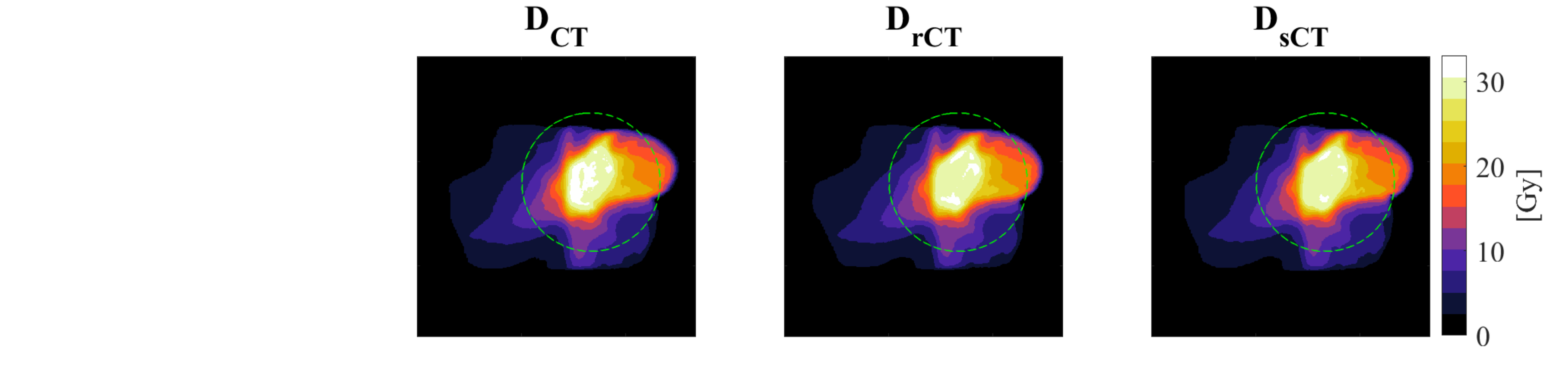}
    \end{subfigure}       \vspace{-25pt}
    \begin{subfigure}{}
     \includegraphics[trim=5 0 10 0,clip,width=0.85\linewidth]{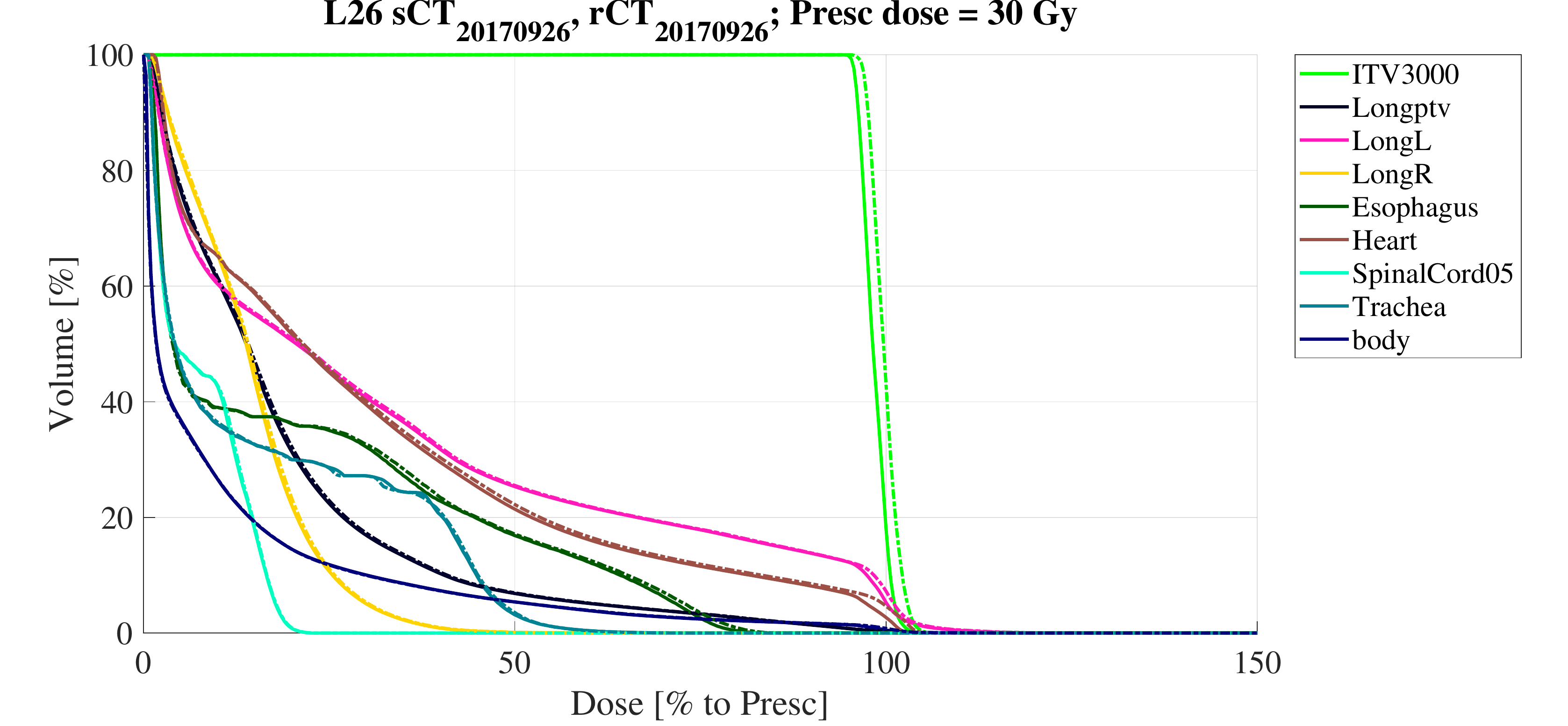} 
    \end{subfigure}          
\end{center}
     \vspace{-20pt}
\captionv{16}{}{Axial views for the lung cancer patient L\thect\ of: (1$^{\textrm{st}}$ row) CBCT (1$^{\textrm{st}}$ column), CT (2$^{\textrm{nd}}$ column), rescan CT (rCT, 3$^{\textrm{rd}}$ column) and synthetic CT (sCT, 4$^{\textrm{th}}$ column), along with (2$^{\textrm{nd}}$ row) the respective difference to rCT, the doses (3$^{\textrm{rd}}$ row). 
 The red, black, or green dotted rectangles indicate the position of Mask$_{\textrm{\tiny CBCT}}$. The days refer to the acquisition date of the rCT. In the 4$^{\textrm{th}}$ row, the DVH is shown for target and OARs of sCT (continuous lines) and rCT (dashed lines).
   \label{fig:Ptexample_Ln}
 }
\end{figure}

Figure~\ref{fig:Ptexample_Hn}, \ref{fig:Ptexample_Br}, \ref{fig:Ptexample_Ln} show examples of CBCT and sCT obtained from the single network for a HN, breast and lung cancer patient, respectively.
One can observe that the network reduced scatter artefacts while retaining anatomical accuracy. Considering an example of lung patient (Figure~\ref{fig:Ptexample_Ln}), one can observe the occurrence of atelectasis between CT and rCT/sCT.

\newcounter{ct}

\setcounter{ct}{24}


\setcounter{ct}{27}


\setcounter{ct}{26}

\subsection{Dose comparison}
Figure~\ref{fig:Ptexample_Hn}, \ref{fig:Ptexample_Br}, \ref{fig:Ptexample_Ln} report also dose distributions calculated on CT, rCT and sCT along with their 
DVHs.
One can qualitatively notice small difference between doses on rCT and sCT. 
When considering the quantitative results, no significant differences were observed between $\textrm{DD}_{\textrm{sCT}}$ and $\textrm{DD}_{\textrm{CT}}$.
On average (Table~\ref{tbl:DoseDiff}), dose differences between sCT/rCT ($\textrm{DD}_{\textrm{sCT}}$) were lower then for CT/rCT ($\textrm{DD}_{\textrm{rCT}}$), e.g.
in the high dose region (D$>90\%$) maximum mean differences in the range [0.1;0.2]\% and [-0.3;0.9]\% were found for $\textrm{DD}_{\textrm{sCT}}$ and $\textrm{DD}_{\textrm{rCT}}$, respectively.

\begin{table}[htbp]
\begin{center}
\captionv{16}{}{Statistics of the dose comparison of the thirty patients in the test set. The values are reported as percentage mean$\pm1\sigma$ and range [min; max].
 }
  \label{tbl:DoseDiff}
\vspace*{2ex}
\small{
\begin{tabular} {|c| c|c|c|  c|c|c| }
\hline
& \multicolumn{3}{c|}{\textbf{sCT vs rCT}}   & \multicolumn{3}{c|}{\textbf{CT vs rCT}} \\  
\hline
\multirow{2}{*}{\textbf{Sites}} & $\textrm{DD}_{\textrm{sCT}}^{1}$ & $\gamma_{3\%,3\,\mathrm{mm}}$ $^{3}$ & $\gamma_{2\%,2\,\mathrm{mm}}$ $^{4}$ &  $\textrm{DD}_{\textrm{rCT}}^{2}$  & $\gamma_{3\%,3\,\mathrm{mm}}$\ $^{3}$ &$\gamma_{2\%,2\,\mathrm{mm}}$ $^{4}$ \\
 & [$\%$] & [$\%$] & [$\%$] & [$\%$] & [$\%$] & [$\%$] \\
 \hline
\multirow{ 2}{*}{\textbf{Head-and-neck}} 
\
& $0.1\pm0.5$ & $98.4\pm1.9$ & $95.3\pm2.8$ &  $0.9\pm1.1$ & $97.0\pm2.3$ & $91.1\pm4.9$ \\ 
& $[-0.8;0.7]$ & \ \ $[93.4;100]$ &\ \ $[90.1;99.6]$ &  $[-1.8;2.0]$ & \ \ $[92.7;99.6]$ & \ \ $[80.7;97.4]$ \\ 
\hline
\multirow{ 2}{*}{\textbf{Breast}} & $0.1\pm0.4$ & $96.5\pm4.1$ & $89.4\pm8.2$ &$-0.3\pm0.8$ & $95.7\pm5.5$ & $88.8\pm9.2$ \\ 
& $[-0.5;0.8]$ & \ \ $[86.1;99.6]$ &\ \ $[70.8;97.8]$ &$[-1.7;0.7]$ & \ \ $[81.1;99.5]$ & \ \ $[66.3;97.5]$ \\
\hline
\multirow{ 2}{*}{\textbf{Lung}} 
& $0.2\pm0.9$ & $97.0\pm2.9$ & $91.3\pm6.2$ &  $-0.1\pm1.5$ & $95.2\pm3.6$ & $87.7\pm6.8$ \\ 
& $[-1.3;1.8]$ & \ \ $[91.5;99.7]$ &\ \ $[82.3;98.4]$ & $[-2.6;3.0]$ & \ \ $[89.8;99.4]$ & \ \ $[79.4;96.6]$ \\
\hline
\end{tabular}} \\
\footnotesize{
$^{c} \textrm{DD}_{\textrm{sCT}}=\frac{\textrm{sCT}-\textrm{rCT}}{\textrm{rCT}} \cdot 100$ on dose $>$ 90\% of the prescribed dose. \\
$^{d} \textrm{DD}_{\textrm{rCT}}=\frac{\textrm{rCT}-\textrm{CT}}{\textrm{CT}} \cdot 100$ on dose $>$90\% of the prescribed dose. \\
$^{e}$Pass rates of $\gamma_{3\%,3\,\mathrm{mm}}$ on dose $>$ 50\% of the prescribed dose.\\
$^{f}$Pass rates of $\gamma_{2\%,2\,\mathrm{mm}}$ on dose $>$ 50\% of the prescribed dose.
}

\end{center}
\end{table}

The mean gamma pass rates with the 2\%,2mm criteria were higher for sCT/rCT compared to CT/rCT for all VOIs, which is in line with the dose differences observed.
All DVH points differed on average $<$0.5\% compared to rCT.
DVH points differences were $<2\%$ except for the heart of a breast patient (B31, -5.6\%), the oesophagus of two breast patients (B30, 3.1\% and B31, 2.3\%), left lung and spinal cord of two lung patients (L25, -2.1\% and L27, 3.8\%, respectively). 
Images of the patients with doses differences in VOIs $>2\%$ were inspected on a single-case basis, as reported in the Supplementary Material. 
We noticed that large dose differences were in low-dose regions, which are more sensitive to statistical differences due to the low amount of events in the Monte Carlo dose calculations.
For a lung case (L25), anatomical differences were reported as the cause of the observed differences.
Also, residual artefacts characterised by inhomogeneous HUs seem to be present along the craniocaudal direction in the lungs for sCT; it appears that for this case the CBCT artefacts were not fully recovered by the network within the lungs.
Also for the other lung case (L27), anatomical differences were observed in the lung. In addition, we noticed the patient was obese and the CBCTs were characterised by severe scatter artefacts.
On sCT, the spinal cord was not entirely recovered, possibly resulting in local difference. Besides, the spinal cord is located in a low-dose region, which may be highlighted when considering metrics as voxel-wise relative differences.

\section{Discussion}
Cycle-consistent generative adversarial network (cycle-GAN) increased the accuracy of HU in CBCT, enabling sCT-based calculations for HN, lung and breast cancer patients.
Also, we found that a single network trained on all the three sites performed similarly to three networks trained on each anatomical site.

When investigating the accuracy of HU on sCT calculating image similarity to rescan CT, we found that HU values were comparable to values observed between CT and rCT. We observed a slight increase in performance for HN compared to breast and lung cancer patients. 
The network was trained with higher amount of slices for HN (1606) compared to lung and breast (1046 and 1016, respectively). We hypothesise that this data imbalancing may have resulted in relatively increased perfomances for HN cancer patients. Also, the use of immobilisation masks for HN case may increase the reproducibility of patient set-up or reduce motion artefacts in the images (both CT and CBCT)~\cite{VanLin2003}.
Though variations in the CBCT imaging protocol were reported, e.g. kV, mAs and linac where the images were acquired, we did not observe any effect on the quality of sCT. It may be of interest to investigate thoroughly the influence to the robustness of the method to variations of acquisition settings, as already proposed by Maier et al.~\cite{Maier2019}. 

In terms of dose calculation accuracy, we compared sCT to rCT, achieving excellent results for all the anatomical sites.
Previous work with deep learning was performed on prostate~\cite{Kida2018,Hansen2018,Landry2019,Kurz2019}, HN~\cite{Liang2019,Li2019} and lung~\cite{Xie2018} patients.
For HN patients, similar findings were reported by Liang et al.~\cite{Liang2019}, where also a cycle-GAN was utilised where a mean ($\pm1\sigma$) 2\%, 2mm $\gamma$ pass-rates of 98.4$\pm$1.7\% was obtained compared to 98.4$\pm1.9$\% of this work. 
Also, Li et al. used a 2D U-net with residual convolutional units achieving mean DVH point difference $<1\%$~\cite{Li2019}.
In our study, similar mean DVH point differences ($<0.5\%$) were achieved, which demonstrates the high sCT quality achieved with our approach.
For lung patients, Xie et al. applied patch-based residual learning on lung patients obtaining a conspicuous correction of cupping and streaking artefacts~\cite{Xie2018}. Unfortunately, they did not perform any dose calculations and used different metrics, making it difficult to compare the studies.

Repositioning inevitably occurred between CBCT, rCT and CT. To further minimise anatomical and set-up differences, we could have recurred to deformable image registration (DIR) to increase the similarity of CBCT/sCT and CT/rCT. However, we opted against it for the following reasons: (i) since we were trying to reproduce the dose derived by CT-based calculations, we did not want to modify CT or rCT further; (ii) residual deformation errors should be thoroughly evaluated~\cite{Paganelli2018}, and this was deemed out of the scope of this investigation; (iii)
recurring to using solely translation mimics the set-up procedure that is currently performed clinically at the linacs, and we aimed at observing the impact of dose evaluation in a comparable setting.

The main limitation of this study is deemed to be the cohort size: ten patients per anatomical sites in the test set may be considered as a low number.
Before clinical implementation, a study including a larger number of patients should be initiated, paying particular attention to the data variability and data balancing among anatomical sites.
Besides, we did not adapt the contours of targets and OARs, which is necessary to investigate the clinical impact of replanning thoroughly.
To our knowledge, this is the most extensive study so far presented utilising a convolutional neural network for sCT generation with ninety-nine included patients. Also, notwithstanding the relatively limited sample, this work offers valuable insights into the generalisation capability of a single cycle-GAN, and, in general, in showing that a single neural network can convert CBCTs of multiple sites.

Currently, we balanced the sites based on the number of patients performing training with about 1.5 times more images for HN compare to lung and breast patients. It would be interesting to investigate in a future study whether different balancing may maintain comparable image similarity for all the anatomical sites.
We believe that further improvements can be made by balancing data in terms of the number of slices included in the training. 

In our study, we showed, for the first time, that a single cycle-GAN can be utilised for multiple anatomical sites as HN, breast and lung. 
This finding has important implications for simplifying the training of a convolutional network since a single network may be adopted for different anatomical sites. 
To fully understand whether a single network may facilitate CBCT-based dose calculations for the whole body, we are currently performing a novel study including additional anatomical areas, e.g. pelvis, lower abdomen and brain. 

The impact of our work is that with a single cycle-GAN CBCTs were converted into CTs, resulting in sCTs that have sufficient quality to enable dose planning. Also, conversion occurred in a matter of seconds, which is line with the sCT generation time reported by other deep learning approaches for lung~\cite{Xie2018}, prostate~\cite{Hansen2018,Landry2019,Kurz2019} and HN~\cite{Li2019,Liang2019}.
We foresee these as an important step toward online ART.
In conventional non-adaptive radiotherapy, this methodology can be used to evaluate the dosimetric impact of anatomical differences occurring during treatment, supporting the decision to perform a rescan CT or not.

In conclusion, a single cycle-GAN was successfully trained to convert
CBCT to CT using unpaired training data of HN, breast and lung cancer patients.
The resulted sCT resembled a diagnostic quality planning CT and featured the anatomy of the daily CBCT.
In terms of dose calculation accuracy, good results were obtained for all the anatomical sites.
In general, the proposed approach enables considerably fast image conversion, and it may facilitate online adaptive radiotherapy treatments.

\section*{Acknowledgements}
\addcontentsline{toc}{section}{Acknowledgements}
We gratefully acknowledge the support of NVIDIA Corporation with the donation of the Titan Xp GPU used for prototyping this research.

\section*{Conflict of Interest Statement}
\addcontentsline{toc}{section}{Conflict of Interest Statement}
M.Maspero, M.H.F. Savenije, T.C.F. van Heijst, J.J.C. Verhoeff, A.N.T. Kotte, A.C. Houweling and C.A.T. van den Berg report no conflict of interests concerning the submitted work.







\section*{References}
\addcontentsline{toc}{section}{\numberline{}References}
\vspace*{-15mm}





\bibliography{./mybibfile.bib}      



\bibliographystyle{./medphy.bst}    


\end{document}